\documentclass[aps,nofootinbib,preprint,tightenlines]{revtex4}

\usepackage{graphicx}
\usepackage{amsmath}
\usepackage{bm} 

\newcommand{\beq}{\begin{eqnarray}}
\newcommand{\eeq}{\end{eqnarray}}
\newcommand{\bqa}{\begin{eqnarray}}
\newcommand{\eqa}{\end{eqnarray}}

\begin{document}

%%%%%%%%%%%%%%%%%%%%%%%%%%%%%%%%%%%%%%%%
%The normal things to start with.
%%%%%%%%%%%%%%%%%%%%%%%%%%%%%%%%%%%%%%%%

\preprint{HISKP-TH-09/33}
\title{Range corrections for two-neutron halo nuclei in effective theory}
\author{David L. Canham}\email{canham@hiskp.uni-bonn.de}
\author{H.-W. Hammer}\email{hammer@hiskp.uni-bonn.de}
\affiliation{Helmholtz-Institut f\"ur Strahlen- und Kernphysik (Theorie)\\
and Bethe Center for Theoretical Physics,
 Universit\"at Bonn, 53115 Bonn, Germany\\}
\date{\today}

%%%%%%%%%%%%%%%%%%%%%%%%%%%%%%%%%%%%%%%%%
%The Abstact
%%%%%%%%%%%%%%%%%%%%%%%%%%%%%%%%%%%%%%%%%

\begin{abstract}
The range corrections to the universal properties and structure of two-neutron halo nuclei
are investigated within an effective quantum mechanics framework.
Treating the nucleus as an effective three-body system,
we make a systematic improvement upon previous calculations by calculating the linear range
corrections at next-to-leading order. 
Since the effective ranges for the neutron-core interactions are not known,
we estimate the effective range to be set by the inverse of the pion mass.
We investigate the possibility of excited Efimov states in two-neutron halo nuclei
and calculate their mean square radii to next-to-leading order.
We find that the effective range corrections are generally small and the leading order predictions 
are very robust.
\end{abstract}

\maketitle

%%%%%%%%%%%%%%%%%%%%%%%%%%%%%%%%%%%%%%%%%%%%%%%

%Brief introduction 
%%%%%%%%%%%%%%%%%%%%%%%%%%%%%%%%%%%%%%%%%%%%%%%
 
\section{Introduction}
\label{sec:Intro}

There is a considerable interest in the universal properties of 
physical systems with large scattering lengths \cite{Bedaque:2002mn,Braaten-05,Epelbaum:2008ga,Platter:2009gz}.
The low-energy behavior of a few-body system with short-range interactions
and scattering length $|a|$ much larger than the range $R$ of the underlying two-body interaction can be described 
in a model-independent way.
This is achieved by formulating an effective theory that takes advantage of the 
separation of scales inherent in the system, thereby using the ratio $R/|a|$ as a small expansion parameter.
The theoretical uncertainty can then be systematically improved by including increasingly 
higher order terms in the expansion. 
These techniques can readily be applied to {\it halo nuclei}: 
a special class of nuclear systems consisting of a tightly bound core and a \lq\lq halo'' 
of lightly bound nucleons \cite{Riisager-94,Zhukov-93,Hansen-95,Jensen-04}. 
There is a growing history of studies which exploit the 
separation of scales found in halo nuclei to describe the system in an effective theory, 
assuming the core to be a structureless particle and treating the nucleus as a few-body 
system of core and valence nucleons 
\cite{Fedorov-94,Amorim-96,Mazumdar-00,Bertulani-02,BHvK2,Yamashita:2004pv,Higa:2008dn,Canham:2008jd}.

In previous work \cite{Canham:2008jd}, we explored
the universal properties and structure of $2n$ halo nuclei
to leading order (LO) in the expansion in $R/|a|$ by describing the halo nuclei as
an effective three-body system consisting of a core and two 
lightly bound valence neutrons. We constructed an effective potential 
made up of two-body contact interactions, and showed that this approach is equivalent 
to an effective field theory (EFT) framework. Our main focus
was the possibility of such three-body systems to display the universal 
Efimov effect \cite{Efimov-70}. 
Assuming the nuclear ground state to be an Efimov state,
we confirmed earlier model studies  showing that $^{20}$C 
is the only $2n$ halo nucleus candidate
to display an excited Efimov state \cite{Fedorov-94,Amorim-96} 
if the $^{19}$C binding energy is less than 165 keV. 
We also calculated the properties of this state.
Moreover, we studied the structure of other known halo nuclei, calculating 
the matter density form factors and mean square radii 
(See also Ref.~\cite{Yamashita:2004pv} for a previous study).

In this paper, we improve upon our earlier calculation by including the 
momentum dependent next-to-leading order (NLO) term in the short-range 
effective potential. While the scattering of two particles at sufficiently low energy is determined 
to leading order by only the scattering length $a$, 
at NLO, the effective range $r_0$ of the two-body interaction comes into play. Including the linear
range correction, we should be able to predict low-energy observables up to 
errors of order $(r_0/a)^2 \sim (M_{low}/M_{high})^2$. 

Effective field theories with contact interactions have already been used to study the NLO range corrections 
to various few-body systems with large scattering lengths, mostly focused on the three-nucleon system. 
The range corrections to S-wave neutron-deuteron scattering 
in the doublet channel were calculated in perturbation theory in Ref.~\cite{Hammer:2001gh}. 
In Refs.~\cite{Bedaque:2002yg,Griesshammer:2004pe,Platter:2006ad}, low-energy three-body observables 
were calculated up to next-to-next-to-leading order (N2LO) using a formalism that partially resums 
higher-order range effects. An analysis of three- and four-body observables to NLO
within the resonating group model was carried out in \cite{Kirscher:2009aj}.
For a more exhaustive list of such higher-order studies in the three-nucleon system, 
see Refs.~\cite{Epelbaum:2008ga,Platter:2009gz} and references therein. 

There has also been interest lately in calculating the effective range corrections to the Efimov 
effect for three identical bosons. Th\o gersen et al. \cite{Thogersen:2008zz} focused 
on the range corrections for ultracold atoms caught in an optical trap.
Platter, Ji, and Phillips have explored the corrections to the Efimov spectrum  
linear in $r_0$ using once-subtracted momentum-space integral equations derived from 
effective field theory \cite{Platter:2008cx}. They showed that discrete scale invariance relates the 
relative corrections for different Efimov states and the bound state spectrum in the 
unitary limit is unaffected by the linear range corrections.

In this work, we focus on the NLO correction to the properties of two-neutron (2$n$) halo nuclei
arising from a non-zero effective range
in the neutron-neutron and neutron-core interactions.
In the next section, we discuss the theoretical framework for this extension.
Our results are presented in Section \ref{sec:HaloNLO}.
We discuss the impact of this correction on the 
possibility of known 2$n$ halo nuclei to have an excited Efimov state and on their binding energies. 
Moreover, we calculate the shift in the mean square radii 
of various halo nuclei due to the non-zero effective range.
Our conclusions are given in Section \ref{sec:Conclusion}.
Finally, some details on the derivation of the Wigner bound and the calculation of the
matter form factors are given in the appendices.

%%%%%%%%%%%%%%%%%%%%%%%%%%%%%%%%%%%%%%%%%%%%%%%
%Theoretical Framework
%%%%%%%%%%%%%%%%%%%%%%%%%%%%%%%%%%%%%%%%%%%%%%%

\section{Theoretical framework}
\label{sec:Equations}

We want to extend the effective quantum mechanics framework used in
Refs.~\cite{Canham:2008jd,Platter:2004qn} to next-to-leading order (NLO). 
This formulation is equivalent to using field-theoretic language for the problem at hand. 
The short-range interactions characteristic of halo nuclei are described using an effective interaction potential.
Using the NLO effective potential, the low-energy behavior of the system will be reproduced 
up to errors proportional to the low-momentum scale $M_{low}$ over the high-momentum scale 
$M_{high}$ squared. At this order, we require two coupling parameters $C_0$ and $C_2$ tuned to reproduce 
the scattering length $a$ and effective range $r_0$ of the two-body interaction. The theory is valid up to a momentum 
scale $M_{high}$ at which the errors become of order one. 

For the large separation of scales involved in halo nuclei, zero-range interactions 
can be used in constructing the effective interaction potential.  
This leads to a separable potential made up of contact interactions 
in a momentum expansion. The two-body S-wave NLO potential can be written as
\beq
\langle \vec{p} \mid V_{eff} \mid \vec{p}'\rangle = g(p) g(p') (C_0 + C_2(p^2 + p'^2) + \ldots\, ),
\label{SwaveVeffNLO}
\eeq
where the dots indicate higher order interactions and $g(p)$ is a regulator function 
(sometimes called the form factor). Of course, the low-energy observables must be 
independent of the regularization scheme, and one can choose the scheme most convenient for calculations. 
For this study, we use a strong cutoff regularization, in which the 
effective potential is set to zero, $V_{eff} = 0$, for momenta $p,p' > \Lambda$. 
This corresponds to the regulator function $g(p)=\theta(\Lambda-p)$ and
allows for a simple inclusion of a three-body force term, as described below. 
A natural choice for the value of $\Lambda$ is $\Lambda \sim M_{high}$, 
but observables are independent of $\Lambda$ after renormalization, up to higher order corrections 
which scale with powers of $1/\Lambda$.

At this point it is worth commenting on the antisymmetry of the full wave function with
respect to the exchange of nucleons in the core and the halo neutrons. In microscopic cluster
models, such as the resonating group model \cite{Wheeler37} or fermionic molecular dynamics \cite{Neff:2003ib},
the wave function is antisymmetric with respect to exchange of all nucleons.
In our halo effective theory the wave function is only antisymmetric with respect to exchange
of the halo nucleons since the nucleons in the core are not resolved.
The error involved in this approximation is governed by the ratio of the
binding momentum of the halo nucleons to the binding momentum of the core, i.e. $M_{low}/M_{high}$. Within the
domain of validity of the effective theory such effects are subsumed in the effective range
parameters of the neutron-core interaction and can be treated in perturbation theory.
This can be understood as follows: The exchange of a nucleon from the core and a halo nucleon
can only give a sizeable contribution if there is a spatial overlap between the wave function of the
core and the wave function of the halo nucleon. The overlap of the wave functions,
however, is determined by the ratio $M_{low}/M_{high}$. If this ratio is of the order 1/3,
one can expect the antisymmetrization effects to be 30\% at leading order and 10\% at next-to-leading
order. For a microscopic study of antisymmetrization effects in the neutron-alpha system, 
see Ref.~\cite{Kamada:2000is}.

The potential (\ref{SwaveVeffNLO}) is then used in the solution of the three-body Faddeev equations in terms of the spectator functions
$F_i(q)$, which represent the dynamics of the core ($i=c$) and the halo neutron ($i=n$). To find 
the bound state of a halo nucleus composed of two valence neutrons and a core with spin zero, the resulting 
coupled integral equations are simply a generalization of the three-boson equation\footnote{In fact, 
the equations are the same for any bound three-body system of two identical particles and a core with spin zero, 
which interact through the pair-wise zero-range potentials given in Eq.~(\ref{SwaveVeffNLO}).} 
(see \cite{Platter:2004qn,canhamPHD} and references therein).  
In the remainder of the paper, we use units such that $\hbar =c = 1$. For convenience, we set
the nucleon mass $m=1$. The resulting integral equations for the spectator functions $F_n(q)$ and  $F_c(q)$
and the bound state energy $B_3 > 0$ are:
\beq
F_n(q) & =& \ {1 \over 2} \int_0^\Lambda dq'q'^2  \bigg[ \left(\tilde{G}_n(q,q';B_3) + {H(\Lambda) \over \Lambda^2} \right) \  t_n(q';B_3) \ F_n(q') 
\nonumber\\
\nonumber\\
& & \ \ \ + \ \left(\tilde{G}_c(q,q';B_3) + {H(\Lambda) \over \Lambda^2} \right) \ t_c(q';B_3) \ F_c(q') \bigg],
\label{Fnstrongfinal}
\eeq
\beq
F_c(q) & =& \ \int_0^\Lambda dq'q'^2 \bigg[ \left(\tilde{G}_c(q',q;B_3) + {H(\Lambda) \over \Lambda^2} \right) \ t_n(q';B_3) \ F_n(q')\bigg].
\label{Fcstrongfinal}
\eeq 
The functions $\tilde{G}_c$  and $\tilde{G}_n$ arise from the free three-body propagator. They are given by
\beq
\tilde{G}_n(q,q';B_3) & = & {A \over q q'} \ln \left({ {2A \over A+1}B_3 + q^2 + q'^2 + {2 \over A+1} qq' \over {2A \over A+1}B_3 + q^2 + q'^2 - {2 \over A+1} qq'} \right),
\label{Gntilde}
\eeq
\beq
\tilde{G}_c(q,q';B_3) & = & {1 \over q q'} \ln \left({ B_3 + q^2 + {A+1 \over 2A}q'^2 +  qq' \over B_3 + q^2 + {A+1 \over 2A}q'^2 - qq'} \right),
\label{Gctilde}
\eeq
where $A$ is the number of nucleons in the core. 

The effects of the interactions are contained in the two-body T-matrices which are obtained by solving the 
Lipp\-mann-Schwinger equation for the neutron-neutron and neu\-tron-core interactions with the
effective potential in Eq.~(\ref{SwaveVeffNLO}). In the kinematics of 
Eqs.~(\ref{Fnstrongfinal}, \ref{Fcstrongfinal}), we have for the neutron-core T-matrix:
\beq
t_n (q';B_3) & = & {A+1 \over \pi A} 
\left( 1 - \tilde{E}_n(q';B_3)\, h_{nc}(\tilde{x}_{nc})\right)
%\nonumber\\\nonumber\\
%		& & \times 
\Bigg[ -{1 \over a_{nc}} f(\tilde{x}_{nc}) + \sqrt{ \tilde{E}_n(q';B_3)}\, {2 \over \pi} \,\arctan(\tilde{x}_n)  \nonumber\\
\nonumber\\
		& &  +  h_{nc}(\tilde{x}_{nc})\,
\tilde{E}_n(q';B_3) \, \left( {2 \over \pi} \Lambda - \sqrt{\tilde{E}_n(q';B_3)}\, {2 \over \pi}\, \arctan(\tilde{x}_n) \right) \Bigg]^{-1},
\label{tnNLO}
\eeq
with
\beq
\tilde{E}_n(q';B_3) = {2A \over A+1}\left(B_3 + {A+2 \over 2(A+1)}q'^2 \right),
\label{En}
\eeq
and for the neutron-neutron T-matrix:
\beq
t_c (q';B_3) & = & {2 \over \pi} \left( 1 - \tilde{E}_c(q';B_3) \, h_{nn}(\tilde{x}_{nn})\right)
%\nonumber\\\nonumber\\
%		& & \times 
\Bigg[ -{1 \over a_{nn}} f(\tilde{x}_{nn}) + \sqrt{ \tilde{E}_c(q';B_3)}\, {2 \over \pi}\, \arctan(\tilde{x}_c) \nonumber\\
\nonumber\\
		& & +  h_{nn}(\tilde{x}_{nn})\,
\tilde{E}_c(q';B_3) \left( {2 \over \pi} \Lambda - \sqrt{\tilde{E}_c(q';B_3)}\, {2 \over \pi}\, \arctan(\tilde{x}_c) \right) \Bigg]^{-1},
\label{tcNLO}
\eeq
with
\beq
\tilde{E}_c(q';B_3) = B_3 + {A+2 \over 4A}q'^2.
\label{Ec}
\eeq
A single index $n$, $c$ indicates the spectator particle and $a_{nn}$, $r_{nn}$, and $a_{nc}$, $r_{nc}$ 
are the $n$-$n$ and $n$-$c$ scattering lengths and effective ranges, respectively. For brevity we 
have defined the variables:
\beq
\tilde{x}_i = {\Lambda \over \sqrt{\tilde{E}_i(q';B_3)}} \qquad \mbox{and} \qquad \tilde{x}_{ni} 
\equiv {\Lambda \over  \sqrt{2\mu E_{ni}}}, \qquad \mbox{where}\quad i=n,c.
\label{xtilde}
\eeq
Here, $E_{ni}$ is the two-body bound (virtual) state energy calculated from the pole of the two-body T-matrix 
at NLO in the limit $\Lambda \rightarrow \infty$, with $\mu$ the corresponding reduced mass. 
In this limit, which implies a cutoff much larger 
than the momentum scales involved in the problem: $\Lambda \gg 1/|a|, \sqrt{ \tilde{E}_i}, \sqrt{E_{ni}}$, 
these T-matrices reproduce the usual effective range expansion at NLO. Therefore, the pole position 
can be found through  the relation:
\beq
E_{ni} = \left(1 - \sqrt{1 - 2r_{ni}/a_{ni}} \right)^2/(2\mu r_{ni}^2).
\label{Eni}
\eeq 
At leading order ($r_{ni} = 0$), this expression reduces to the universal formula for the two-body binding energy 
for systems with a large positive scattering length: $E_{ni} = 1/(2\mu a_{ni}^2)$. Moreover, we have simplified
the expressions for the two-body T-matrices~(\ref{tnNLO}, \ref{tcNLO}) by defining the
functions 
\beq
f(\tilde{x}) = {2 \over \pi} \left(\arctan\left(\tilde{x}\right) + {1 \over \tilde{x}}\right),
\label{regfunc}
\eeq
which approaches one as $\tilde{x}$ becomes large and
\beq
h_{ni}(\tilde{x}_{ni})&=& {1\over \Lambda^2} {2  - \pi f(\tilde{x}_{ni})\Lambda r_{ni}/2
\over 2 - \pi f(\tilde{x}_{ni})/(\Lambda a_{ni})},\quad
i=n,c,
\label{regfunc2}
\eeq
which vanishes for large $\tilde{x}_{ni}$. The term proportional to $h_{ni}(\tilde{x}_{ni})\, \Lambda$
generates the effective range term in the two-body T-matrices (\ref{tnNLO}, \ref{tcNLO}).

Finally, the dimensionless function $H(\Lambda)$ is a short-range three-body force. This term was included
in the integral equations (\ref{Fnstrongfinal}, \ref{Fcstrongfinal}) 
in analogy to the three-boson case which our equations reproduce
in the limit of identical particles. It corresponds to a three-body contact interaction which is
required for consistent renormalization~\cite{Bedaque:1998kg,Platter:2004qn}.
For a given value of $\Lambda$, the three-body term $H$ can be tuned to reproduce a given 
three-body observable and then other low-energy observables can be predicted using the same value 
$H$. The running of the three-body force $H(\Lambda)$ with the cutoff $\Lambda$ is governed by a limit
cycle. As a consequence, it is always possible to find a value of the 
cutoff $\Lambda$ at which the three-body force vanishes. Hence, at leading
order $\Lambda$ can simply be used as the three-body parameter
in practical calculations \cite{Hammer:2000nf}.
The correct renormalization in the effective theory was explicitly verified
\cite{Platter:2004qn}.
Apart from the three-body observable,
our input parameters at NLO are the $n$-$c$ and $n$-$n$ scattering lengths (or energies) and 
the corresponding effective ranges.
The three-body binding energies are then given by the values of $B_3$ for which the coupled integral equations 
(\ref{Fnstrongfinal}, \ref{Fcstrongfinal}) have a nontrivial solution.

For most halo nuclei, the $n$-$c$ scattering length is not as well known as the two-body bound (virtual) state energy. 
Therefore, we will generally use the two-body energy $E_{nc}$ as an input parameter.  
From $E_{nc}$ and $r_{nc}$, we can calculate the corresponding $a_{nc}$ 
to be used in the two-body T-matrices through Eq.~(\ref{Eni}). 
The $n$-$c$ effective ranges for most halo nuclei are not known experimentally. 
However, the range of the interaction between two nucleons is determined by the exchange of pions. 
We therefore use the inverse of the pion mass, 
$m_{\pi} = 140$ MeV, to estimate the $n$-$c$ effective range, $r_{nc} \approx 1/m_{\pi} = 1.4$ fm. This is equivalent 
to setting the effective range to the natural low-energy length scale of the system.
The value of the $n$-$n$ effective range is known experimentally: $r_{nn} = (2.75 \pm 0.11)$ fm \cite{Miller:1990iz}. 
Its size is comparable to the inverse of the pion mass scale. In our numerical calculations we  
use  $r_{nn}\approx r_{nc}  \approx 1/m_{\pi} = 1.4$ fm. This choice is sufficient for
our purpose of estimating the size of the 
range corrections and testing the robustness of the leading order results.
Moreover, it allows a slightly larger range of cutoffs to be consistent with the Wigner bound.

The Wigner bound is based on general arguments and
constrains the value of the effective range that can be generated by short-range potentials.
It has been shown that the renormalization of an effective potential made up of contact interactions 
in a momentum expansion can only be performed if certain constraints are 
placed on the effective range \cite{Phillips:1997xu}. More specifically, the effective range must be 
negative in the zero-range limit of the potential.
In our case, this  corresponds to the limit $\Lambda\to \infty$ since a finite cutoff generates a finite range.
This constraint follows directly from a general bound on the derivatives of the phase shifts derived first by 
Wigner in 1954 \cite{Wigner:1955zz}. Wigner derived this bound from the fundamental principles of causality and 
unitarity, holding that a scattered wave cannot leave the scatterer before the incident wave has reached it.  
Phillips and Cohen derived the relation between 
the Wigner bound on the derivatives of phase shifts with the constraint on the effective range 
for short-range potentials \cite{Phillips:1996ae}.
This bound applies even if the potential does not vanish exactly outside some finite radius but
merely decreases fast enough for the wave function to approach the asymptotic solution sufficiently 
quickly. The Wigner bound for zero-range potentials was shown to hold true no matter how many terms in the momentum 
expansion are included in the potential \cite{Phillips:1997xu}.
A very recent study has generalized the Wigner Bound to arbitrary dimension and angular momentum \cite{Hammer:2009zh}. 
An explicit derivation of the Wigner bound constraint for our effective potential is given in Appendix \ref{wignerapp}.
In our numerical results in the next section, the cutoff is always chosen in accordance with the Wigner bound.

\section{Effective Range Corrections to 2\boldmath$n$ Halo Nuclei}
\label{sec:HaloNLO}

In this section we present numerical results for the effective range corrections
to low-energy observables of 
three-body halo nuclei composed of a core and two valence neutrons. 
We begin with the range corrections to the possibility of the 
Efimov effect in 2$n$ halo nuclei, as well as to the binding energy of a possible Efimov excited state in $^{20}$C. 
After that, we estimate and discuss effective range corrections to the mean square radii results. 

%%%%%%%%%%%%%%%%%%%%%%%%%%%%%%%%%%%%%%%%%%%%%%%%%%
%NLO Corrections to Halo Nuclei Binding Energies
%%%%%%%%%%%%%%%%%%%%%%%%%%%%%%%%%%%%%%%%%%%%%%%%%%
\subsection{Range Corrections to 2\boldmath$n$ Halo Nuclei Binding Energies}
\label{sec:Carbon20NLO}

In Ref.~\cite{Canham:2008jd}, we have calculated boundary curves 
representing the existence of an excited Efimov state for various values of the core mass in 
the parametric region defined by the ratios $(E_{nc}/B_3^{(n)})^{1/2}$ versus $(E_{nn}/B_3^{(n)})^{1/2}$, 
where $E_{nc}$ and $E_{nn}$ are the $n$-$c$ and $n$-$n$ two-body energies related to the S-wave scattering length to LO. 
An analogous study was carried out previously in Ref.~\cite{Amorim-96} 
within the renormalized zero-range model and motivated our investigation.
It should be noted, however, that this type of analysis implicitly assumes that the ground state of a halo
nucleus is an Efimov state. More general scenarios where only the excited state has universal character
cannot be excluded. 
The curve itself is built up of the points for which the $B_3^{(n+1)}$ binding energy is 
equal to the scattering threshold. This is equivalent to finding the critical scattering lengths 
of the three-body system. 
In similar studies looking at the range corrections to the Efimov effect in the three-boson system 
\cite{Platter:2008cx,Thogersen:2008zz}, it was observed that including a non-zero effective range 
shifted the positions of the critical scattering lengths. 
However, for all values of effective ranges $r_{nc}$ and $r_{nn}$ within the range of validity of
our effective theory ($r_{ni} \ll |a_{ni}|$, $i=n,c$),
we find these shifts to be extremely small such that the boundary curves calculated with a 
non-zero effective range are indistiguishable from those at LO (See Fig.~2 of Ref.~\cite{Canham:2008jd}). 

As a consequence, the main conclusion from \cite{Canham:2008jd} is only slightly changed by the
NLO range corrections: if the $n$-$^{18}$C bound state energy satisfies $E_{nc} <161$ keV, 
$^{20}$C will have an excited Efimov state.
%The error in the three-body ground state energy of
%$^{20}$C is small compared to $E_{nc}$. 
In the following, we investigate the dependence on the $n$-$^{18}$C bound state energy in more detail.
In Ref.~\cite{Canham:2008jd}, we calculated the value of the possible 
excited state energies at LO as a function of $E_{nc}$, using the standard value for 
$a_{nn} = (-18.7 \pm 0.6)$ fm \cite{Gonzales-99}, 
and fixing the cutoff $\Lambda$ 
to reproduce the experimental value of the ground state energy 
$B_3^{(0)} = 3506$ keV \cite{TUNL,Audi-95}. 
Now using the estimate $r_0=r_{nn} = r_{nc} \approx 1/m_{\pi} = 1.4$ fm
for the effective ranges along with the standard value for $a_{nn}$ given above, we calculate 
the excited state energies $B_3^{(1)}$ as a function of $E_{nc}$ to NLO, by fixing $\Lambda$ and tuning 
the three-body term $H$ to reproduce the experimental 
value of the ground state energy $B_3^{(0)} = 3506$ keV. 
For our calculations we use the cutoff $\Lambda = 580$ MeV. 
%%%%%%%%%%%%%%%%%%%%%%%%%%%%%%%%%%%%%%%%%%%%%%%%%%%%%%%%%%
\begin{figure}[t]
	\centerline{\includegraphics*[width=10cm,angle=0]{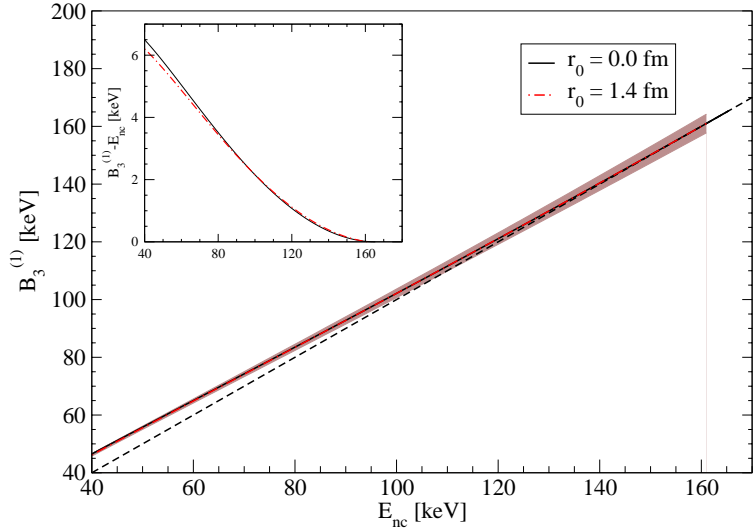}}
	\caption{Binding energy of the $^{20}$C excited Efimov state as a function of the $n$-$^{18}$C bound state energy 
		to LO (solid black line) and NLO (dotted-dashed red line) with NLO error bands. 
		The dashed line represents the scattering threshold which is given by
    $B_3^{(1)}= E_{nc}$. The inset shows the excited state energy relative to the scattering threshold. }
	\label{fig:20CNLO}
\end{figure}
%%%%%%%%%%%%%%%%%%%%%%%%%%%%%%%%%%%%%%%%%%%%%%%%%%%%%%%%%%
The NLO result is compared to the LO result in Fig.~\ref{fig:20CNLO}, 
where the solid black line is the excited state energy to LO, the dotted-dashed red line to NLO, and 
the dashed line represents the scattering threshold. The inset
graph shows the excited state energy relative to the scattering threshold.
Whereas the excited state in the LO calculation exists when $E_{nc} < 165$ keV, the state to NLO only 
exists when $E_{nc} < 161$ keV. For larger values of $E_{nc}$, the $^{20}$C system moves 
across the scattering threshold, and the excited Efimov state disappears. 
We also see that the effective range corrections only lead to a recognizable shift 
toward smaller binding energies for $E_{nc} < 90$ keV. For larger values of $E_{nc}$ 
there is a region where the effective range corrections lead to a shift towards larger binding energies. 
However, this shift is always very small, no greater than $\approx 0.5$ keV 
over the whole range of possible $E_{nc}$ values shown, and always smaller than the NLO error bands. 

We have estimated this NLO error using the theoretical uncertainty of our effective potential. 
The uncertainty in binding energies calculated using the two-body effective potential of 
Eq.~(\ref{SwaveVeffNLO}) is $\approx (r_0/a)^2$. For effective ranges much smaller than the scattering 
length we know that $1/a_{nc}^2 \approx 2\mu_{nc} E_{nc}$. 
Therefore, we estimate the theoretical uncertainty in the binding energy 
of the excited state using the sum
of the uncertainties from the $n$-$n$ and $n$-$^{18}$C interactions: 
$\Delta B_3^{(1)}/ B_3^{(1)}\approx(r_{nn}/a_{nn})^2 + r_{nc}^2 (2\mu_{nc} E_{nc})$. 

The TUNL nuclear data evaluation from 1995 \cite{TUNL,Audi-95} quotes the
$n$-$^{18}$C bound state energy as $E_{nc} = (162 \pm 112)$ keV. This value is
based on experiments using strong breakup reactions \cite{Viera86,Gillebert87,Wouters88,Orr91}
and would leave the possibility of an Efimov excited state in $^{20}$C open.
The more
recent atomic mass evaluation AME2003~\cite{AME}, however, quotes
$E_{nc} = (580 \pm 100)$ keV which excludes an excited Efimov state.
This evaluation includes a recent measurement of $E_{nc}$ using Coulomb
dissociation of $^{19}$C that found $E_{nc} = (530 \pm 130)$ keV~\cite{Nakamura-99}.
Typel and Baur showed  that the value of Ref.~\cite{Nakamura-99} is also consistent the 
distribution of the $B(E1)$ continuum strength \cite{Typel04}. A review of the experiments was 
given in \cite{Aumann05}.

%We note that there is a {\bf more recent} value of the $n$-$^{18}$C bound state energy \cite{Nakamura-99},
%$E_{nc} = (530 \pm 130)$ keV, which {\bf is supported by the atomic mass evaluation found in Ref.~\cite{AME}.
%Unfortunately, the magnitude of this binding energy excludes an excited Efimov state in $^{20}$C.}
%This value was obtained from the angular distributions in Coulomb breakup of $^{19}$C. Typel and Baur showed 
%that the $n$-$^{18}$C energy essentially fixes the distribution of the $B(E1)$ continuum strength \cite{Typel04}.
%A discussion of this issue from an experimental point of view was given in \cite{Aumann05}. The discrepancy between
%the two {\bf different} values for $n$-$^{18}$C energy is particularly relevant for the question of excited Efimov states
%in $^{20}$C and deserves further study. {\bf However, this issue is outside the scope of this work.}

%%%%%%%%%%%%%%%%%%%%%%%%%%%%%%%%%%%%%%%%%%%%%%%%%%
%NLO Corrections to Halo Nuclei Radii
%%%%%%%%%%%%%%%%%%%%%%%%%%%%%%%%%%%%%%%%%%%%%%%%%%
\subsection{Range Corrections to 2\boldmath$n$ Halo Nuclei Mean Square Radii}
\label{sec:HaloradiiNLO}

In this subsection, we will calculate the NLO corrections to the mean square radii 
of 2$n$ halo nuclei. We will briefly review the necessary formalism and then present our results. 
A more detailed description of the formalism is given in Refs.~\cite{Canham:2008jd,canhamPHD}.
For the calculation of the one- and two-body matter density form factors 
and mean square radii, we need the full three-body
wave function, which can be reconstructed from the Faddeev spectator functions $F_n$ and $F_c$.
The relevant formulae are given in Appendix \ref{sec:EigtoWF}. 

The mean square radii are then calculated from the 
matter density form factors in the low momentum transfer region, where the slope of the 
form factor determines the mean square radius $\left\langle r^2 \right\rangle$: 
\beq
{\mathcal F}(k^2) & = & 1 - {1 \over 6} k^2 \left\langle r^2 \right\rangle + \ldots\, .
\label{FFexpandHaloNLO}
\eeq
Of course, the mean square radius acquired depends on the choice of one- or two-body form factor. 
These matter density form factors are calculated from the full wave function of 
the three-body system through Eqs.~(\ref{FFiNLO}, \ref{FFncNLO}, \ref{FFnnNLO}).
The slope of the two-body form factor ${\mathcal F}_{ni}(k^2)$, where $i=n,c$, 
will give the mean square distance between the two 
particles in the chosen two-body subsystem, either $\left\langle r_{nn}^2 \right\rangle$ or 
$\left\langle r_{nc}^2 \right\rangle$. Analogously, the 
slope of the one-body form factor ${\mathcal F}_{i}(k^2)$ will give the mean square distance of the spectator 
particle from the center of mass of the two-body subsystem, 
either $\left\langle r_{c-nn}^2 \right\rangle$ or $\left\langle r_{n-nc}^2 \right\rangle$. 
However, it is more 
useful to calculate the distance of the individual particles from the center of mass of the three-body bound 
state. If $b_i$ is the slope of the one-body form factor ${\mathcal F}_{i}(k^2)$ at $k^2 = 0$, 
the mean square radius of one 
of the bodies $i$ from the three-body center of mass is given by:
\beq
\left\langle r_i^2 \right\rangle = -6b_i \left(1 - {m_i \over 2m_n + m_c} \right)^2,
\label{rfromb1HaloNLO}
\eeq
where $m_i$ is the mass of the desired constituent, $i=n,c$, and $m_n$ and $m_c$ are the neutron and core masses, 
respectively. 
%The various radii of the three-body system are illustrated in Fig.~\ref{fig:Haloradii}.
%%%%%%%%%%%%%%%%%%%%%%%%%%%%%%%%%%%%%%
%\begin{figure}[t]
%	\centerline{\includegraphics*[width=5cm,angle=0]{radifig.eps}}
%	\caption{The various radii of the three-body system.}
%	\label{fig:Haloradii}
%\end{figure}
%%%%%%%%%%%%%%%%%%%%%%%%%%%%%%%%%%%%%%%%

We have extracted the radii by fitting a polynomial in $k^2$ to the form factor results  
for small $k^2$. We have used polynomials of varying degree up to 5th order in $k^2$ in order to verify 
the stability and convergence of the fit.
We have found a satisfactory stability in the slope when fitting to a polynomial 
to the fourth order in $k^2$, up to a value of $k^2$ at which the form factor has dropped less than 10 percent.

The extracted radii for known halo nuclei to both LO and NLO are shown in Table~\ref{table:radiiNLO}. 
\begin{table*}[tb]
\begin{small}
	\centerline{\begin{tabular}{||c|c|c|c||c|c|c|c||} \hline \hline
	Nucleus & $B_3$ [keV] & $E_{nc}$ [keV] & $r_0$ [fm] & $\sqrt{\left\langle r_{nn}^2 \right\rangle}$ [fm] & 
        $\sqrt{\left\langle r_{nc}^2 \right\rangle}$ [fm] & $\sqrt{\left\langle r_{n}^2 \right\rangle}$ [fm] & 
        $\sqrt{\left\langle r_{c}^2 \right\rangle}$ [fm] \\ \hline 
	 	$^{11}$Li & 247 & -25 & 0.0 & 8.7$\pm$0.7 & 7.1$\pm$0.5 & 6.5$\pm$0.5 & 1.0$\pm$0.1 \\
	  & 247 & -25 & 1.4 & 8.80$\pm$0.07 & 7.21$\pm$0.06 & 6.51$\pm$0.05 & 1.040$\pm$0.008 \\
	 	& 247 & -800~\cite{Wilcox-75} & 0.0 & 6.8$\pm$1.8 & 5.9$\pm$1.5 & 5.3$\pm$1.4 & 0.9$\pm$0.2 \\
	 	& 247 & -800~\cite{Wilcox-75} & 1.4 & 6.3$\pm$0.5 & 5.5$\pm$0.4 & 4.9$\pm$0.4 & 0.81$\pm$0.06 \\ \hline
 	 $^{14}$Be & 1120 & -200 \cite{Thoennessen-00} & 0.0 & 4.1$\pm$0.5 & 3.5$\pm$0.5 & 3.2$\pm$0.4 & 0.40$\pm$0.05 \\ 
 	  & 1120 & -200 \cite{Thoennessen-00} & 1.4 & 3.86$\pm$0.09 & 3.29$\pm$0.08 & 3.02$\pm$0.07 & 0.384$\pm$0.009 \\ \hline 
	$^{12}$Be & 3673 & 503 & 0.0 & 3.0$\pm$0.6 & 2.5$\pm$0.5 & 2.3$\pm$0.5 & 0.32$\pm$0.07 \\ 
	  & 3673 & 503 & 1.4 & 3.3$\pm$0.2 & 2.7$\pm$0.1 & 2.5$\pm$0.1 & 0.35$\pm$0.02 \\ \hline
 	 $^{18}$C & 4940 & 731 & 0.0 & 2.6$\pm$0.7 & 2.2$\pm$0.6 & 2.1$\pm$0.5 & 0.18$\pm$0.05 \\ 
 	  & 4940 & 731 & 1.4 & 2.9$\pm$0.2 & 2.4$\pm$0.2 & 2.3$\pm$0.2 & 0.21$\pm$0.01 \\ \hline
 	 $^{20}$C & 3506 & 530 \cite{Nakamura-99}  & 0.0 & 3.0$\pm$0.7 & 2.5$\pm$0.6 & 2.4$\pm$0.5 & 0.19$\pm$0.04 \\
  	& 3506 & 530 \cite{Nakamura-99}  & 1.4 & 3.38$\pm$0.18 & 2.75$\pm$0.15 & 2.60$\pm$0.14 & 0.21$\pm$0.01 \\
  	& 3506 & 162 & 0.0 & 2.8$\pm$0.3 & 2.4$\pm$0.3 & 2.3$\pm$0.3 & 0.19$\pm$0.02 \\ 
 	  & 3506 & 162 & 1.4 & 3.03$\pm$0.06 & 2.53$\pm$0.05 & 2.39$\pm$0.05 & 0.198$\pm$0.004 \\ 
  	& 3506 & 60 & 0.0 & 2.8$\pm$0.2 & 2.3$\pm$0.2 & 2.2$\pm$0.2 & 0.18$\pm$0.01 \\
  	& 3506 & 60 & 1.4 & 2.84$\pm$0.03 & 2.41$\pm$0.03 & 2.28$\pm$0.03 & 0.192$\pm$0.002 \\ 
  	$^{20}$C* & 65.0$\pm$6.8 & 60 & 0.0 & 42$\pm$3 & 38$\pm$3 & 41$\pm$3 & 2.2$\pm$0.2 \\ 
    $^{20}$C* & 64.9$\pm$0.7 & 60 & 1.4 & 43.2$\pm$0.5 & 38.7$\pm$0.4 & 42.9$\pm$0.5 & 2.26$\pm$0.02 \\ \hline \hline
	\end{tabular}}
	\end{small}
\caption{\label{table:radiiNLO}
        Various mean square radii of different halo nuclei. The second two columns show the input values for 
        the three-body ground state energy and the two-body $n$-$c$ energy (negative values corresponding to 
        virtual energies), respectively, as given by \cite{TUNL}, except where otherwise noted. The fourth 
        column shows the input value for both two-body effective ranges, related to LO ($r_0 = 0.0$ fm) 
        or NLO ($r_0 = 1.4$ fm) calculations. 
        The rows marked by $^{20}$C* show the results 
				for the excited Efimov state of $^{20}$C, with binding energy displayed in the second column, 
				which is found above the ground state ($B_3=3506$ keV).}
\end{table*}
We show only a selection of the LO results found in Ref.~\cite{Canham:2008jd}, to give a general 
overview of NLO corrections to the LO results.  As input we have used the standard 
value of the $n$-$n$ scattering length, $a_{nn} = (-18.7 \pm 0.6)$ fm \cite{Gonzales-99}, 
along with the experimental values of the $n$-$c$ two-body energies $E_{nc}$ 
shown in the third column of Table~\ref{table:radiiNLO} (negative values correspond to virtual energies). 
The effective range of both two-body subsystems is shown in the fourth column, indicating which results refer to 
LO ($r_0=0$ fm) or NLO ($r_0=1.4$ fm)
calculations. As before, we use the inverse of the pion mass to estimate the effective range 
of both the $n$-$n$ and the $n$-$c$ interactions for all 2$n$ halo nuclei. 
As a three-body input, the three-body term is tuned to reproduce the experimental ground state 
binding energy $B_3^{(0)}$ shown in the second column of Table~\ref{table:radiiNLO}.  
These experimental values for the two-body and three-body energies are taken from the most recent results of the 
"Nuclear Data Evaluation Project" of TUNL \cite{TUNL}, except where otherwise noted. 
The calculations are performed at a fixed cutoff $\Lambda$. Due to the Wigner bound, 
some care must be taken in choosing $\Lambda$. 
For the range  of scattering lengths, or corresponding two-body energies, which we wish to explore,
the Wigner bound leads to a maximum value for the 
cutoff for most of the 2$n$ halo nuclei of approximately $\Lambda < 630$ MeV. 
For our calculations, we use a cutoff $\Lambda = 580$ MeV for all nuclei. This value is significantly larger
than the typical momenta in halo nuclei which are well below the pion mass.

The theoretical error for the NLO calculation is again estimated by the uncertainty of the two-body effective potential, 
Eq.~(\ref{SwaveVeffNLO}), which is of order $(r_0/a)^2$, where $r_0$ is the effective range of the interaction, 
and $a$ the scattering length. 
The uncertainty in the radii is then calculated exactly as was done in the previous subsection, 
using the sum of the uncertainties from the $n$-$n$ and $n$-$c$ interactions: 
$\Delta \langle r^2 \rangle/\langle r^2\rangle\approx  (r_{nn}/a_{nn})^2 + r_{nc}^2 (2\mu_{nc} E_{nc})$. 
We stress that our NLO results as well as their errors are based on our estimate of the effective range, $r_0=1.4$ fm.
They do not include the uncertainty in the estimate of $r_0$ and therefore must be interpreted with some care.

\begin{figure*}[t]
	\centerline{\includegraphics*[width=12cm,angle=0]{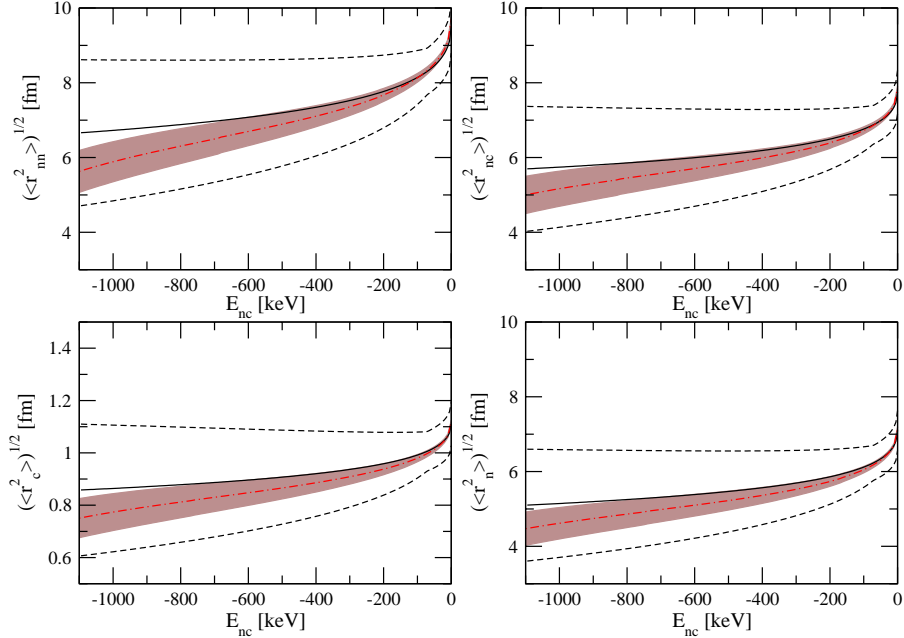}}
	\caption{The various mean square radii for $^{11}$Li as a function of the $n$-$^{9}$Li two-body energy 
	with error bands from the theoretical uncertainty. The LO results represented by the solid black lines, with error 
	bands represented by dashed lines (as calculated in Ref.~\cite{Canham:2008jd}). 
	The NLO results represented by the red dotted-dashed lines with solid error bands.
	As input, the $n$-$n$ scattering length $a_{nn} = -18.7$ fm, the three-body binding energy $B_3^{(0)} = 247$ keV, 
	and for NLO results the effective range $r_0 = 1.4$ fm were used.}
	\label{fig:11LiRvsEnc0NLO}
\end{figure*}

We will now discuss the results for the NLO corrections to the mean square radii due to the 
effective range of the interactions.
For the {\it Borromean} halo nuclei $^{11}$Li and $^{14}$Be, in which all two-body subsystems are unbound, 
the general tendency is for a positive effective range to shift all mean square radii to smaller values. 
The only exception is the case of $^{11}$Li using the central value of the $n$-$c$ energy 
$E_{nc} = (-25\pm 15)$ keV \cite{TUNL}.  This difference is due to the fact that this 
value of the virtual energy is very close to the resonant limit, $E_{nc} = 0.0$, while the competing value 
$E_{nc}=(-800\pm 250)$ keV \cite{Wilcox-75} is much larger in comparison. This can be seen more 
clearly if we look at the mean square radii over a range of $E_{nc}$ values. Using the central value of the 
three-body binding energy as input, $B_3^{(0)} = 247$ keV, the NLO results are plotted in comparison 
to the LO values in Fig.~\ref{fig:11LiRvsEnc0NLO}, with error bands estimated from the theoretical 
uncertainty, as described above. 
%%%%%%%%%%%%%%%%%%%%%%%%%%%%%%%%%%%%%%
%%%%%%%%%%%%%%%%%%%%%%%%%%%%%%%%%%%%%%%%
The solid black lines represent the LO results, with error bands represented by the dashed lines~\cite{Canham:2008jd}. 
The dotted-dashed red lines represent the NLO results, with solid colored error bands. 
We see that the mean square radii to NLO show the same behavior with changing $E_{nc}$ as the LO case, 
with a decrease in the magnitude of $E_{nc}$ leading to an increase in size, with a more rapid increase 
as the resonant limit is approached. However, for the NLO results, the increase is more rapid when 
$E_{nc}$ approaches zero, such that below some value of $|E_{nc}|$, a positive effective range 
shifts the mean square radii to larger values. Otherwise, a positive effective range always shifts 
the mean square radii to smaller values.  One can see that this shift is more drastic as the magnitude of $|E_{nc}|$ 
grows larger.  This is most likely due to the fact that for these large values of $|E_{nc}|$, 
the scattering length becomes smaller, nearing the order of magnitude of the effective range. 
For example, the virtual $n$-$^{9}$Li energy $E_{nc} = -1000$ keV, 
along with the effective range $r_{nc} = 1.4$ fm, corresponds to a scattering length 
$a_{nc} = -4.2$ fm. 

The study by Marqu\'es et al. \cite{Marques-00} extracted experimental values for 
$\sqrt{\left\langle r_{nn}^2 \right\rangle}$ for Borromean halo nuclei using the technique of 
intensity interferometry along with the two neutron correlation function 
to study the dissociation at intermediate energies of two neutrons in halo nuclei. 
The result for $^{11}$Li, $\sqrt{\left\langle r_{nn}^2 \right\rangle}_{exp} = (6.6\pm 1.5)$ fm, 
is in close agreement with both the LO, $\sqrt{\left\langle r_{nn}^2 \right\rangle} = (6.8\pm 1.8)$ fm, 
and NLO, $\sqrt{\left\langle r_{nn}^2 \right\rangle} = (6.3\pm 0.5)$ fm, calculations found using 
the two-body virtual energy reported in \cite{Wilcox-75}, $E_{nc} = -800$ keV. However, 
due to the large uncertainties in both the theoretical and experimental values 
for $^{11}$Li, there exists a large range of $E_{nc}$ values which would produce a 
$\sqrt{\left\langle r_{nn}^2 \right\rangle}$ value in agreement with the experimental value of 
Marqu\'es et al.~\cite{Marques-00}. The LO result for $^{14}$Be, 
$\sqrt{\left\langle r_{nn}^2 \right\rangle} = (4.1\pm 0.5)$ fm, is about a Fermi below the 
the experimental value from Ref.~\cite{Marques-00},  
$\sqrt{\left\langle r_{nn}^2 \right\rangle}_{exp} = (5.4\pm 1.0)$ fm, but the two values are
consistent within one error bar. The NLO result, 
$\sqrt{\left\langle r_{nn}^2 \right\rangle} = (3.86\pm 0.09)$ fm, is even smaller
and only consistent with the experimental value within two sigma errors. 
Recall, however, that the NLO 
results are found using only an estimate for the effective ranges. Our results indicate that the effective
range in the $n$-$^{12}$Be-interaction is below this estimate.
Furthermore, the large uncertainty in these experimental values is 
indicative of the need for more precise measurements of the mean square distances in 2$n$ halo nuclei. 
Also, the recent work by Orr \cite{Orr:2008px} discusses the care which must be taken in interpreting 
the experimental results of \cite{Marques-00}. Specifically, the technique is sensitive to the population 
of states in the continuum by the dissociation process rather than being a true ground state measurement.  
For more details on this issue, see Ref.~\cite{Orr:2008px}.

%%%%%%%%%%%%%%%%%%%%%%%%%%%%%%%%%%%%%%
\begin{figure*}[t]
	\centerline{\includegraphics*[width=12cm,angle=0]{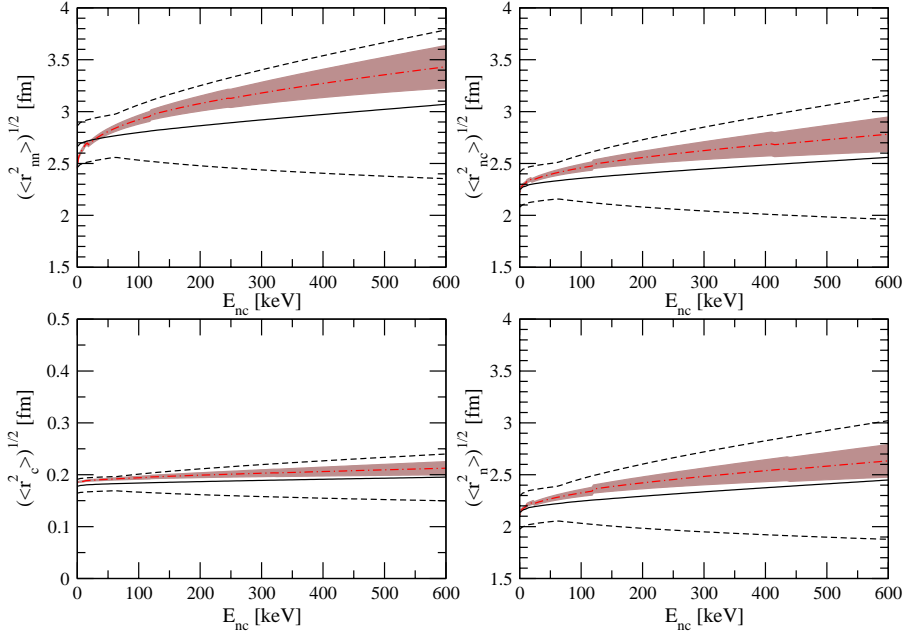}}
	\caption{The various mean square radii for $^{20}$C as a function of the $n$-$^{18}$C two-body energy 
	with error bands from the theoretical uncertainty. The LO results represented by the solid black lines, with error 
	bands represented by dashed lines (as calculated in Ref.~\cite{Canham:2008jd}). 
	The NLO results represented by the red dotted-dashed lines with solid error bands.
	As input, the $n$-$n$ scattering length $a_{nn} = -18.7$ fm, the three-body binding energy $B_3^{(0)} = 3506$ keV, 
	and for NLO results the effective range $r_0 = 1.4$ fm were used.}
	\label{fig:20CRvsEnc0NLO}
\end{figure*}
%%%%%%%%%%%%%%%%%%%%%%%%%%%%%%%%%%%%%%%%
Turning now to the so called {\it Samba} halo nuclei $^{12}$Be, $^{18}$C, and $^{20}$C, 
for which the $n$-$c$ subsystem is bound, it can be seen that 
a positive effective range always shifts the mean square radii to larger values. 
We focus on the case of $^{20}$C
%, as the large uncertainty in the $n$-$c$ energy,
%with {\bf the already mentioned two different experimental} values, 
%$E_{nc} = (162 \pm 112)$ keV \cite{TUNL}, and $E_{nc} = (530 \pm 130)$ keV \cite{Nakamura-99}, 
%suggests that we 
and look at the mean square radii over a range of $E_{nc}$ values.  
The results, using the central value for the 
three-body binding energy as input, $B_3^{(0)} = 3506$ keV, were calculated to LO in 
Ref.~\cite{Canham:2008jd}. We have now calculated these values to NLO, using the 
inverse of the pion mass as an estimate of the effective range, and the results compared to the LO case 
can be seen in Fig.~\ref{fig:20CRvsEnc0NLO}, with error bands 
estimated from the theoretical uncertainty, as described above. 
The solid black lines represent the LO results, with error bands represented by the dashed lines. 
The dotted-dashed red lines represent the NLO results, with solid colored error bands. 
We see that the mean square radii to NLO show the same behavior with changing $E_{nc}$ as the LO case, 
with an increase in $E_{nc}$ leading to an increase in size. Also we see more clearly the general 
shift toward larger radii due to the effective range corrections. 

%%%%%%%%%%%%%%%%%%%%%%%%%%%%%%%%%%%%%%
\begin{figure*}[t]
	\centerline{\includegraphics*[width=12cm,angle=0]{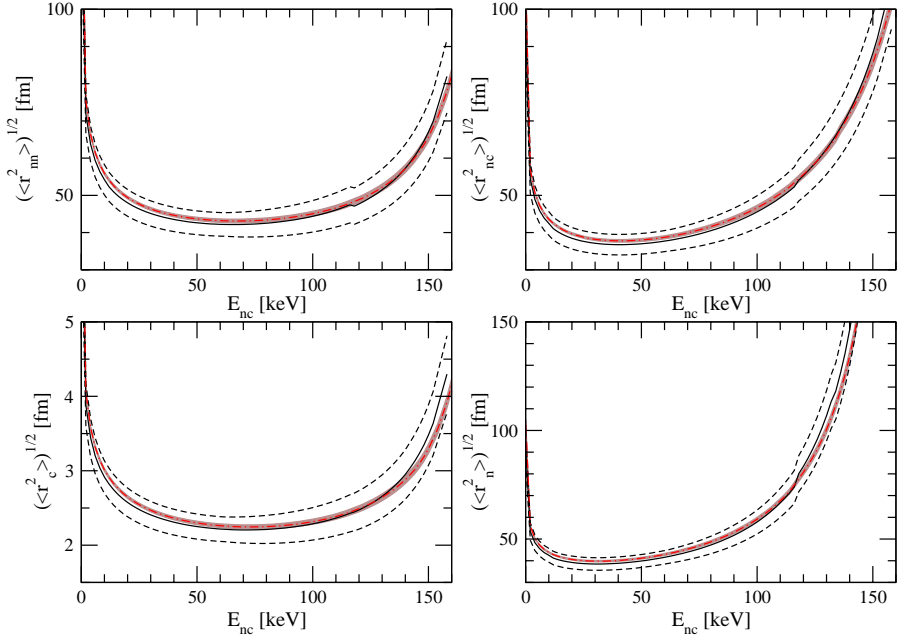}}
	\caption{The various mean square radii for the Efimov excited state of $^{20}$C as a function of the 
	$n$-$^{18}$C two-body energy with error bands from the theoretical uncertainty. 
	The LO results represented by the solid black lines, with error 
	bands represented by dashed lines (as calculated in Ref.~\cite{Canham:2008jd}). 
	The NLO results represented by the red dotted-dashed lines with solid error bands.
	As input, the $n$-$n$ scattering length $a_{nn} = -18.7$ fm, the three-body binding energy $B_3^{(0)} = 3506$ keV, 
	and for NLO results the effective range $r_0 = 1.4$ fm were used.}
	\label{fig:20CRvsEnc1NLO}
\end{figure*}
%%%%%%%%%%%%%%%%%%%%%%%%%%%%%%%%%%%%%%%%
As was shown in the previous subsection, there exists one Efimov excited state in $^{20}$C 
for $E_{nc} < 165$ keV. 
In Fig.~\ref{fig:20CRvsEnc1NLO}, the mean square radii for this excited state to NLO are plotted over a 
range of $E_{nc}$ values, together with the LO results from Ref.~\cite{Canham:2008jd}. 
Again we see that the general behavior of the mean square radii as a function of the two-body energy 
remains the same when calculated to NLO. At the endpoints, signifying the critical $E_{nc}$ where 
the excited state breaks up, the mean square radii diverge as the excited state is destroyed and the 
particles fly apart. For small values of $E_{nc}$, there is a positive shift in the 
mean square radii for positive effective range, although this shift is relatively small, as compared to the relative 
shift in the ground state radii. However, as $E_{nc}$ grows, there is a region where a positive effective 
range causes a negative shift in the mean square radii. This is related to the region where 
the effective range causes a positive shift in the binding energy, as described in 
the previous subsection. 

In Table~\ref{table:radiiNLO}, we have highlighted the results using two different values of the 
$n$-$c$ two-body binding energy: $E_{nc} = (530 \pm 130)$ keV \cite{Nakamura-99} and the older value
$E_{nc} = (162 \pm 112)$ keV \cite{TUNL}.
Moreover, we show results for $E_{nc} = 60.0$ keV which would lead to an Efimov excited state. 
The rows marked by $^{20}$C$^*$ give the results for the excited state. The three-body binding energy 
of this excited state is listed in the second column with theoretical uncertainty as 
calculated in Sec.~\ref{sec:Carbon20NLO}.

Within the error estimates our results generally agree with the values obtained previously by Yamashita et al. using the
renormalized zero-range model \cite{Yamashita:2004pv}. Note, however, that this model does not
include the effective range corrections and no error estimates are given.

Finally, we note that we explicitly do not apply our method to $^6$He. 
The $n$-$^4$He-interaction is dominated by strong P-wave resonance in the $J=3/2$ channel, 
while our effective theory assumes that S-wave interactions dominate. A treatment
of $^6$He therefore requires an effective theory with a different counting scheme. 
While various schemes to treat such P-wave resonances in Effective Field Theory have been developed
\cite{Bertulani-02,BHvK2}, their application to three-body systems remains to be worked out.

%%%%%%%%%%%%%%%%%%%%%%%%%%%%%%%%%%%%%%%%%%%%%%%%
%Conclusion
%%%%%%%%%%%%%%%%%%%%%%%%%%%%%%%%%%%%%%%%%%%%%%%%
\section{Conclusion}
\label{sec:Conclusion}

In this paper, we have studied the next-to-leading order (NLO) effective range corrections to the 
universal properties of two-neutron halo nuclei. Our investigation 
was performed within an effective quantum mechanics framework using cutoffs of the order 500 MeV. 
Assuming that the halo nuclei have resonant S-wave interactions between the neutron and the core, 
the effective potential reduces to a separable S-wave potential.
At NLO the potential is determined by the neutron-neutron and neutron-core energies and effective
ranges plus one three-body parameter that can be fixed from any three-body datum.
At  next-to-next-to-leading order in the expansion in $M_{low}/M_{high}$ no new two-body
parameters enter. Whether another three-body parameter enters at this order is still an open question 
that has not been completely resolved~\cite{Bedaque:2002yg,Platter:2006ad}. 

Our main focus was in investigating NLO corrections to the results for 2$n$ halo nuclei which arise from a 
non-zero effective range. Because the neutron-core effective range is not known experimentally, we were only
able to estimate the range corrections assuming the effective range to be of the order of the inverse 
pion mass,  $r_0 \approx 1 / m_{\pi} = 1.4$ fm. Based on this assumption, we found that the leading order results
are very robust and range corrections are typically small.
We investigated the possibility of excited Efimov states in halo nuclei assuming that the ground state
is also an Efimov state.
In particular, we found that the corrections to the parametric boundary curves, 
within which at least one excited Efimov state will occur were essentially unchanged for realistic values of the 
effective range. 
%Assuming the ground state to be an Efimov state {\bf and using the experimental 
%binding energy for $^{19}$C reported in Ref.~\cite{Audi-95}}, the only halo candidate to display an 
%excited Efimov state remains $^{20}$C. 
We found that the value of the critical $n$-$^{18}$C two-body energy required for such an excited state to exist
in $^{20}$C was decreased slightly from $165$ keV at LO to $161$ keV at NLO.
While the TUNL nuclear data evaluation from 1995 \cite{TUNL,Audi-95} quotes this energy as $E_{nc} = (162 \pm 112)$ keV,
the most recent evaluation AME2003~\cite{AME} finds $E_{nc} = (580 \pm 100)$ keV which excludes an excited Efimov state in $^{20}$C.
We also calculated the NLO shift in the excited state binding energy, finding that $|\Delta B_3^{(1)}|<0.5$ keV, 
a nearly negligible shift from the LO result.

Moreover, we calculated the shift from the LO results in the mean square radii for known 2$n$ halo nuclei. 
We found that for {\it Borromean} halo nuclei, the general tendency is for a positive effective range to shift 
all mean square radii to smaller values, unless the two-body $n$-$c$ virtual energy is very close to threshold. 
The opposite was found to be true for the so called {\it Samba} halo nuclei, for which the $n$-$c$ subsystem 
is bound. The positive effective range shifts the mean square radii of these halo nuclei to larger values. 
We also compared our results for the $n$-$n$ mean square radius to the experimental data for the Borromean halo nuclei
$^{11}$Li and $^{14}$Be \cite{Marques-00}. 
While there is good agreement in the $^{11}$Li system, the results for $^{14}$Be only agree within two sigma errors. 
This indicates that the effective range in the $n$-$^{12}$Be-interaction is smaller than our naive estimate $r_0=1.4$ fm.
%This, along with the large error bars in both experimental and theoretical results,
%indicates that a better analysis could be done if more accurate measurements are available.  
For a better description within our effective theory, it will be necessary to determine the as yet unknown parameters, 
specifically the various neutron-core scattering lengths and effective ranges. 
%With new improved experimental data for these weakly bound nuclei, 
%much more knowledge can be obtained about the properties of halo nuclei. 
Such information will also provide a 
solid basis for evaluating their universal behavior and determining the relevance of the Efimov effect
as a possible binding mechanism for halo nuclei.

If the Efimov effect is responsible for the binding of at least some three-body halo nuclei 
or their excited states
it would open the possibility of a universal binding mechanism for higher-body halo nuclei as well.
In Ref.~\cite{Hammer:2006ct}, a pair of universal tetramer states associated with every Efimov trimer was 
predicted for the case of identical bosons and their binding energies were calculated in the vicinity
of the unitary limit. Recently, the spectrum of these states was mapped out
for all scattering lengths and their experimental signature in 4-body recombination was pointed out
\cite{Stecher:2008}. Soon after this prediction, the loss resonances from both tetramers in an
ultracold gas of $^{133}$Cs atoms were indeed observed \cite{Ferlaino:2009}.
Another calculation indicated that
even higher-body cluster states with universal properties might exist \cite{Stecher09}, leading to an
intriguing paradigm for a universal binding mechanism of weakly bound nuclei near the drip lines. Effective
theories represent an ideal tool to explore this scenario in nuclei and predict its signatures in the structure
and reactions of halo nuclei. Together with new precise
experimental data from facilities such as FAIR and FRIB this will open the possibility to test this
scenario.
We note that this scenario does not require that the ground state of a halo
nucleus is an Efimov state. More general situations where only excited states have universal character
are also possible.

\begin{acknowledgments}

We thank Gerhard Baur, Vitaly Efimov, Andreas Nogga, and Daniel Phillips for discussions.
This research was supported in part by the DFG through
SFB/TR 16 \lq\lq Subnuclear structure of matter'' and by the BMBF
under contracts No. 06BN411 and No. 06BN9006. 

\end{acknowledgments}

%%%%%%%%%%%%%%%%%%%%%%%%%%%%%%%%%%%%%%%%%%%%%%%%
%Appendix: F_i to WF to FF
%%%%%%%%%%%%%%%%%%%%%%%%%%%%%%%%%%%%%%%%%%%%%%%%
\appendix

\section{Wigner Bound}
\label{wignerapp}
In this appendix, we derive the Wigner bound on the S-wave effective range for our effective potential. 
In the renormalization process, the bare coupling constants must be redefined twice in order to 
absorb high-energy effects proportional to positive powers of the cutoff parameter, which is equivalent to adding 
counter terms to the effective potential. The bare couplings $\{C_0,C_2\}$ of Eq.~(\ref{SwaveVeffNLO}) 
are related to the redefined couplings $\{\tilde{C}_0,\tilde{C_2}\}$ by:
\beq
C_0 = {\tilde{C}_0 \over 1 - 4\pi^2 2\mu \tilde{C_2} {2 \over \pi} {\Lambda^3 \over 3}} + {2 \over \pi} {\Lambda^5 \over 5} 2\pi^2 2\mu C_2^2,
\label{C0toC01R2}
\eeq
and 
\beq
C_2 = {1 \over 2\pi^2 2\mu {2 \over \pi} {\Lambda^3 \over 3}} \left( -1 + \sqrt{{1 \over 1 - 4\pi^2 2\mu \tilde{C_2} {2 \over \pi} {\Lambda^3 \over 3}}} \right),
\label{C1toC1R2}
\eeq
where $\mu$ is the two-body reduced mass. 
One clearly sees that there are real values of $\tilde{C_2}$ for which the bare couplings become complex. This 
would imply a complex effective potential. Therefore, we must place a constraint on the value of $\tilde{C_2}$:
\beq
4\pi^2 2\mu \tilde{C_2} {2 \over \pi} {\Lambda^3 \over 3} & \leq & 1.
\label{C1R2constraint}
\eeq

The renormalization of the effective potential is completed by tuning the redefined coupling constants 
to reproduce the scattering length $a$ and the effective range $r_0$ according to the following 
expressions:
\beq
{1 \over \tilde{C_0}} = 2\pi^2 2\mu \left[ {1 \over a} f(\tilde{x}_0) - {2 \over \pi} \Lambda \right],
\label{C0toaNLO}
\eeq
and
\beq
\tilde{C_2} = {1 \over 4\pi^2 2\mu} \left[ {{r_0 \over 2} f(\tilde{x}_0) - {2 \over \pi} {1 \over \Lambda} \over \left( {1 \over a} f(\tilde{x}_0) - {2 \over \pi} \Lambda \right)^2 } \right],
\label{C1tor0a0}
\eeq
where the parameter $\tilde{x}_0$ is analogous to $\tilde{x}_{ni}$ from Eq.~(\ref{xtilde}) found with 
$a_{ni}=a$ and $r_{ni}=r_0$. Substituting Eq.~(\ref{C1tor0a0}) into Eq.~(\ref{C1R2constraint}), 
we find the Wigner bound for our effective potential:
\beq
{2 \over \pi} {\Lambda^3 \over 3} {{r_0 \over 2} f(\tilde{x}_0) - {2 \over \pi} {1 \over \Lambda} \over \left( {1 \over a} f(\tilde{x}_0) - {2 \over \pi} \Lambda \right)^2 } \leq 1.
\label{WignerBound}
\eeq
If the cutoff is much larger than the momentum scale of the T-matrix pole, $\Lambda \gg \sqrt{2\mu E_2}$, then 
$f(\tilde{x}_0) \rightarrow 1$ 
and we have:
\beq
r_0 & \leq & 2 \left( {2 \over \pi}{4 \over \Lambda}  - {6 \over a \Lambda^2} + {\pi \over 2}{3 \over a^2\Lambda^3}  \right).
\label{WignerBoundSlimit}
\eeq
It is then obvious that for $\Lambda\to\infty$,
the effective range must be equal to or less than zero.  Conversely, by setting the input values of the 
scattering length and effective range, Eq.~(\ref{WignerBound}) places a constraint 
on the maximum value of the cutoff $\Lambda$. 

\section{Wave function and matter density form factors}
\label{sec:EigtoWF}

The full bound state wave function of a two-neutron halo nucleus can be reconstructed from the solutions for the 
spectator functions $F_n$ and $F_c$ found from the coupled integral 
equations~(\ref{Fnstrongfinal}, \ref{Fcstrongfinal}). The form of the wave function, 
$_i\langle pq|\Psi \rangle \equiv \Psi_i(p,q)$, depends on the choice of two-body 
subsystem and corresponding spectator particle, where the index $i=n,c$ 
labels the spectator particle. Recall that in the wave functions, the Jacobi momentum $p$ 
describes the relative momentum between the two particles in the chosen two-body subsystem, 
while $q$ describes the momentum of the spectator particle relative to the center of mass of the two-body subsystem.

To reconstruct the S-wave part of the full wave functions with either a neutron or core spectator from 
the spectator functions we find: 
\beq
\Psi_n(p,q) & = & \left( G^n_0(p,q;B_3) + {H(\Lambda) \over \Lambda^2} \right) \bigg[ t_n(q;B_3) F_n(q)  \nonumber\\
\nonumber\\
	& &  + \ {1 \over 2} \int_{-1}^{1} dx \ t_n(\tilde{\pi}'_{nn};B_3) F_n(\tilde{\pi}'_{nn}) \ + \ t_c(\tilde{\pi}'_{nc};B_3) F_c(\tilde{\pi}'_{nc}) \bigg], 
\label{PsinFstrong} \\ \nonumber\\
\Psi_c(p,q) & = & \left( G^c_0(p,q;B_3) + {H(\Lambda) \over \Lambda^2} \right) \bigg[ t_c(q;B_3) F_c(q)  \nonumber\\
\nonumber\\
	& & \, \,  + \ \int_{-1}^{1} dx \ t_n(\tilde{\pi}'_{cn};B_3) F_n(\tilde{\pi}'_{cn}) \bigg],
\label{PsicFstrong}
\eeq
where the two-body T-matrices $t_n$ and $t_c$ are given in Eqs.~(\ref{tnNLO}, \ref{tcNLO}). 
The expressions for the propagators $G_0^n$ and $G_0^c$ are:
\beq
G_0^n(p,q;B_3) &=& \left[B_3 + {A+1 \over 2A}p^2 + {A+2 \over 2(A+1)}q^2\right]^{-1},
\label{propn}\\
G_0^{c}(p,q;B_3) &=& \left[B_3 + p^2 + {A+2 \over 4A}q^2\right]^{-1}\,.
\label{propc}
\eeq
Finally, the shifted momenta are: 
\beq
\tilde{\pi}'_{nn} &\equiv& \tilde{\pi}'_{nn}(p,q) = \sqrt{p^2 + {1 \over (A+1)^2}\, q^2 - {1\over A+1}\,2 pqx},
\label{pitilde'nnpq}
\eeq
\beq
\tilde{\pi}'_{nc} &\equiv& \tilde{\pi}'_{nc}(p,q) = \sqrt{p^2 + {A^2 \over (A+1)^2}\, q^2 - {A\over A+1} 2pqx},
\label{pitilde'ncpq}
\eeq
and
\beq
\tilde{\pi}'_{cn} &\equiv& \tilde{\pi}'_{cn}(p,q) = \sqrt{p^2 + {1 \over 4} q^2 - pqx}.
\label{pitilde'cnpq}
\eeq

The three-body wave functions can be used to calculate other low-energy 
properties of the three-body bound state. With the Jacobi momentum states it is straightforward
to calculate the Fourier transform of the one- and two-body matter densities with respect to the momentum 
transfer squared. These are defined as the one- and two-body matter density form factors 
${\mathcal F}_{i}(k^2)$ and ${\mathcal F}_{ni}(k^2)$, respectively, where $i = n,c$. The derivation of 
the form factors from the three-body S-wave wave functions can be found in Refs.~\cite{Canham:2008jd,canhamPHD}.  
The resulting expression for the one-body form factors is:
\beq
{\mathcal F}_i(k^2) & = & \int dp \ p^2 \int dq \ q^2 \int_{-1}^1 dx \ \Psi_i(p,q) \Psi_i(p,\sqrt{q^2+k^2-2qkx}), \ 
\label{FFiNLO}
\eeq
while the two-body form factors are:
\beq
{\mathcal F}_{nc}(k^2) & = & \int dp \ p^2 \int dq \ q^2 \int_{-1}^1 dx \ \Psi_n(p,q) \Psi_n(\sqrt{p^2+k^2-2pkx},q), \ 
\label{FFncNLO}
\eeq
and
\beq
{\mathcal F}_{nn}(k^2) & = & \int dp \ p^2 \int dq \ q^2 \int_{-1}^1 dx \ \Psi_c(p,q) \Psi_c(\sqrt{p^2+k^2-2pkx},q). \ 
\label{FFnnNLO}
\eeq

%%%%%%%%%%%%%%%%%%%%%%%%%%%%%%%%%%%%%%%%%%%%%%%%
%Bibliography
%%%%%%%%%%%%%%%%%%%%%%%%%%%%%%%%%%%%%%%%%%%%%%%%

\end{document}